# A SCALABLE VIDEO SEARCH ENGINE BASED ON AUDIO CONTENT INDEXING AND TOPIC SEGMENTATION

Julien LawTo[1], Jean-Luc Gauvain[2], Lori Lamel[2], Gregory Grefenstette[1], Guillaume Gravier[3], Julien Despres[4], Camille Guinaudeau[3], Pascale Sebillot[3]

[1]Dassault Systèmes/Exalead, Paris, France; [2]LIMSI, Orsay, France ; [3]IRISA, Rennes, France ; [4]Vocapia Research, Orsay, France

E-mail: [1]3ds.com, [2]limsi.fr, [3]irisa.fr, [4]vocapia.com

*Abstract:* **One important class of online videos is that of news broadcasts. Most news organisations provide near-immediate access to topical news broadcasts over the Internet, through RSS streams or podcasts. Until lately, technology has not made it possible for a user to automatically go to the smaller parts, within a longer broadcast, that might interest them. Recent advances in both speech recognition systems and natural language processing have led to a number of robust tools that allow us to provide users with quicker, more focussed access to relevant segments of one or more news broadcast videos. Here we present our new interface for browsing or searching news broadcasts (video/audio) that exploits these new language processing tools to (i) provide immediate access to topical passages within news broadcasts, (ii) browse news broadcasts by events as well as by people, places and organisations, (iii) perform cross lingual search of news broadcasts, (iv) search for news through a map interface, (v) browse news by trending topics, and (vi) see automatically-generated textual clues for news segments, before listening. Our publicly searchable demonstrator currently indexes daily broadcast news content from 50 sources in English, French, Chinese, Arabic, Spanish, Dutch and Russian.**

## 1    INTRODUCTION

Many of the major news organisations provide immediate access to topical news broadcasts over the internet, through RSS streams or podcasts. In parallel, many users rely on third-party sites [1] to describe topical extracts of longer news broadcasts. However, in spite of early attempts on broadcast news retrieval and browsing from speech [1, 2, 3], technology has not made it possible for a user to efficiently find small segments of interest from longer broadcasts within a large collection spanning multiple languages. In particular, work on topic segmentation of broadcast news, e.g. [4, 5] was limited in the number of shows and languages that could be handled.

Most current video search engines rely, in a large part, on indexing the textual metadata associated with the video (title, tags, surrounding page-text). Videos that are returned for a search over common search engines are those which contain those search terms in their metadata.

Recent progress in spoken language processing (in transcription, topic segmentation, keyword extraction) has led to a number of robust tools that allow us to now provide users with quicker and more focussed access to relevant segments of one or more news broadcast videos. Researchers have focused on either one or the other aspect of the processing chain. For example, new approaches to topic segmentation of broadcast news have been proposed in [6, 7, 8], the latter focusing on robustness to different shows and to transcription errors. Still, few systems integrate all of these components in a complete and comprehensive large scale demonstration able to return portions of videos relevant to a query, while also providing query-free browsing capabilities.

This paper presents a complete system demonstrating an alternative approach for browsing and searching videos and audio newscasts based on robust spoken document processing in multiple languages. In broadcast news, most of the linguistic information is encoded in the audio channel of video data, which, once transcribed, can be processed using natural language processing and semantic processing techniques. The interface presented here integrates many of these technologies to provide topical access to automatically identified broadcast segments.

The main novelty here, with respect to similar available systems (including previous versions of our own) is the topical segmentation of news broadcasts. All the search and browsing tools are generated from automatically detected topic segments, using methods described below. Search is constrained to each segment, and these segments are returned as the result of a search, though the entire broadcast is still accessible, if desired. For example, if a user searches *for Ron Paul AND Barack Obama*, only those segments in which both politicians are mentioned will be returned. Our interface also provides additional, more elaborate optional annotations for each segment: named entities, timestamps of mentions of each query term, a pincushion timeline bar showing mentions, and for each segment, a label of corpus-derived important terms mentioned in that segment. Querying can be performed in a language different from broadcast language, exploiting commonly available translation tools.

---

[1] For example, reddit.com, huffingtonpost.com, newser.com, ...

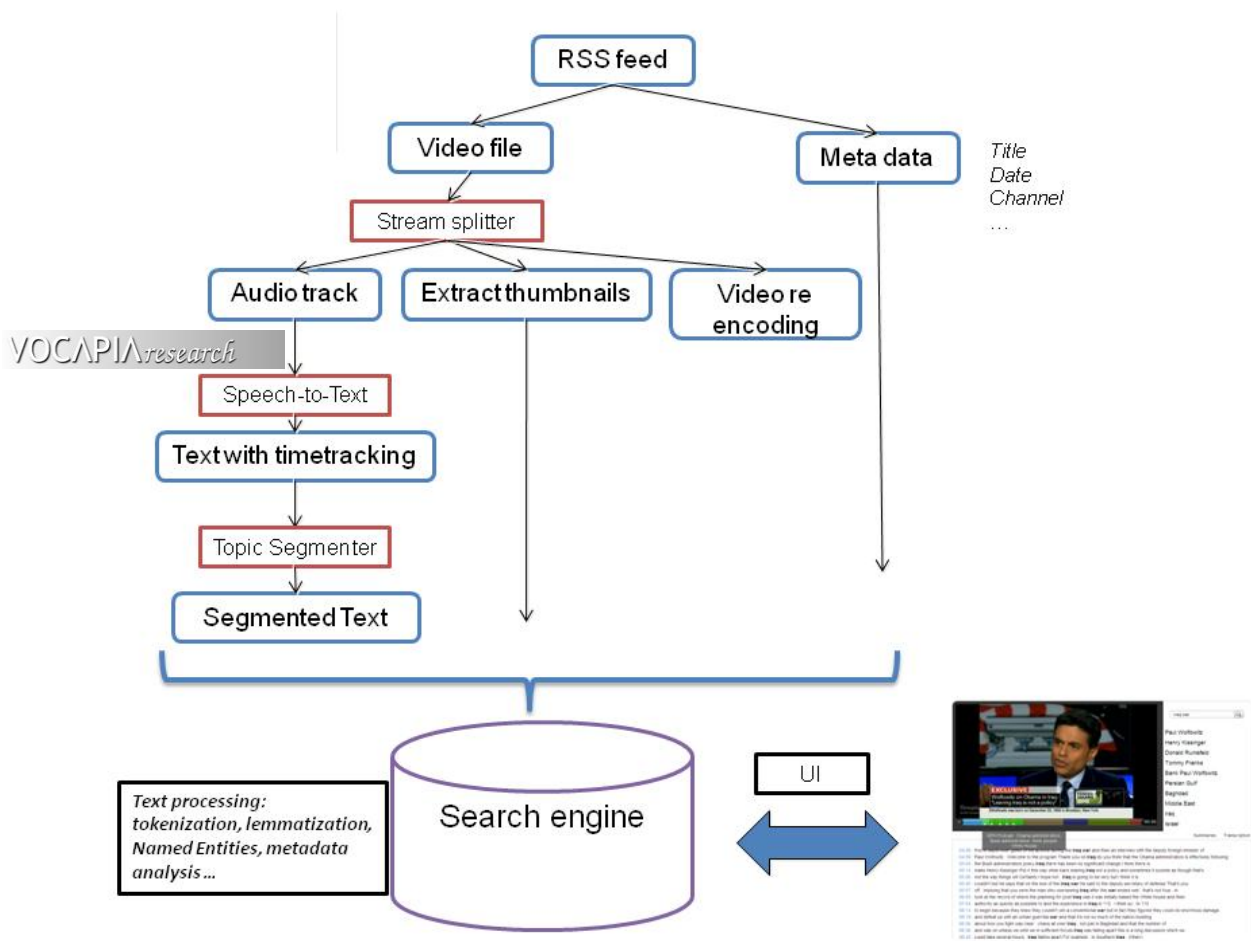

**Figure 1 Overview of the architecture of the video indexing**

The paper is organized as follows: Back-end processing of the video and audio sources is described in Section 2. Section 3 describes the user interface, illustrating the various features of the system. In Section 4, processing time is discussed, followed by a conclusion and a description of future evolution of our system.

## 2 BACK END: OFF LINE PROCESSING

Our system indexes freely available podcast sources, broadcast via RSS streams. Figure 1 presents the main steps of the processing of these podcasts.

### 2.1 Automatic Speech Recognition (ASR)

State-of-the art speech transcription systems for 7 languages (French, English, Spanish, Mandarin, Dutch, Russian and Arabic) are the core of the demonstration. The transcription systems make use of statistical modelling techniques similar to those described in [9, 10], which gives details for an English broadcast news system. The acoustic and language models and pronunciation dictionaries are language dependent [11, 12], and trained on large audio and text corpora. Speech decoding is carried out in a single pass with statistical n-gram language models, and takes less time than the signal duration. Proper case is output for all languages, and postprocessing converts numerical quantities for amounts, dates, telephone numbers to Arabic numbers for English, French, and Spanish. The system outputs an xml file containing the words identified in the audio document, along with their time codes and a transcription confidence measure.

Our first processing step partitions the data into speech segments, and, after determining the gender, clusters segments from the same speaker. This information can later be combined with the content in the automatic transcription to associate true speaker names to parts of the data. Each language has recognition word lists containing from 50k to 300k words which generally give a good coverage of the language. However, "breaking news" may have repeated occurrences of words that are unknown to the system. New functionality has recently been incorporated which allows users to update the recognition word list and this technology is currently undergoing experimentation. This automatic speech recognition technology used in our system has been frequently demonstrated to obtain top performance in international benchmarks.

### 2.2 Topic segmentation

A news broadcast is often divided into stories, which may have no relation with each other. If the broadcast is transcribed into one textual document, a complex search, such as *Barack Obama in China,* may return videos in which China is mentioned in one story and Barack Obama in another, contrary to what the user intended to find. To remedy this problem in the newest version of our system, we process the uninterrupted textual output of the automatic speech transcription by applying topic segmentation to break the transcript of a show into topically homogeneous segments. These segments would ideally correspond to individual reports in classical news.

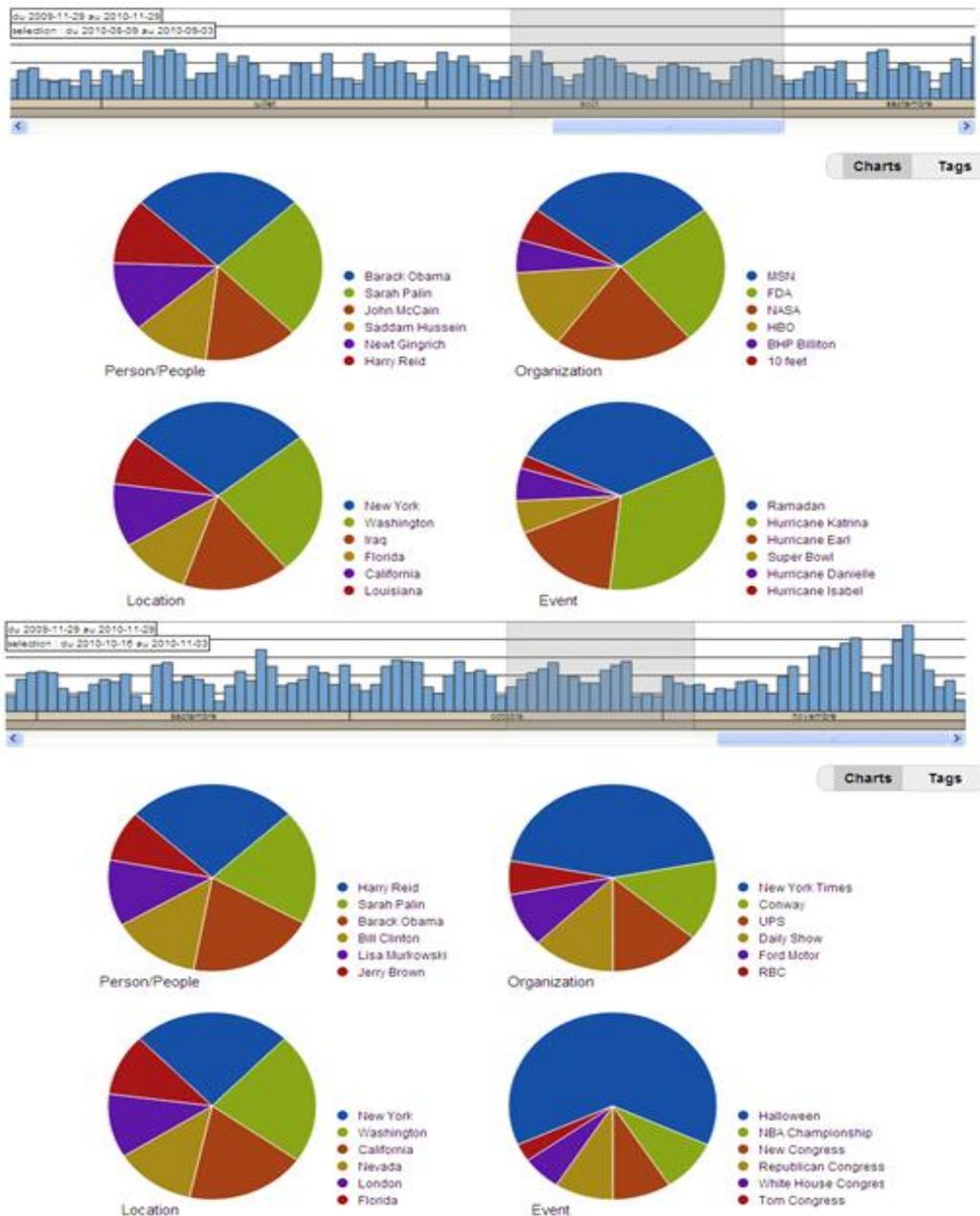

**Figure 2 News trends over different periods**

Topic segmentation has been studied in natural language processing since the early 1990s [13]. Most approaches use vocabulary differences in windows over the document to detect subject changes and topic shifts. Our system relies on an extension of the linear segmentation methods described in [14]. The general idea of this lexical cohesion based method is to search for the best possible segmentation among all the possible ones. A generalized probability criterion is used to measure thematic cohesion of a segment, exploiting repetitions in the vocabulary: A unigram language model, estimated from the word counts in the segment, is used to compute the probability of the word sequence corresponding to the segment. Our language model estimation has been improved with respect to [14]. We added features to account for the peculiarities of broadcast news transcripts, namely transcription errors and the limited number of repeated words due to stylistic reasons. In particular, word level confidence measures are used to deal with transcription errors while semantic relations are introduced to counteract the limited number of repetitions with the same methods as in [8].

In practice, each word in the transcript is labelled with part-of-speech tags and lemmatized. Computation of thematic coherence probability is limited to nouns, adjectives and non-modal verbs. The output of the topic segmentation process is a set of segments. As a by-product, that we exploit, each segment can be labelled by the few keywords which most significantly contributed to the lexical cohesion of the segment.

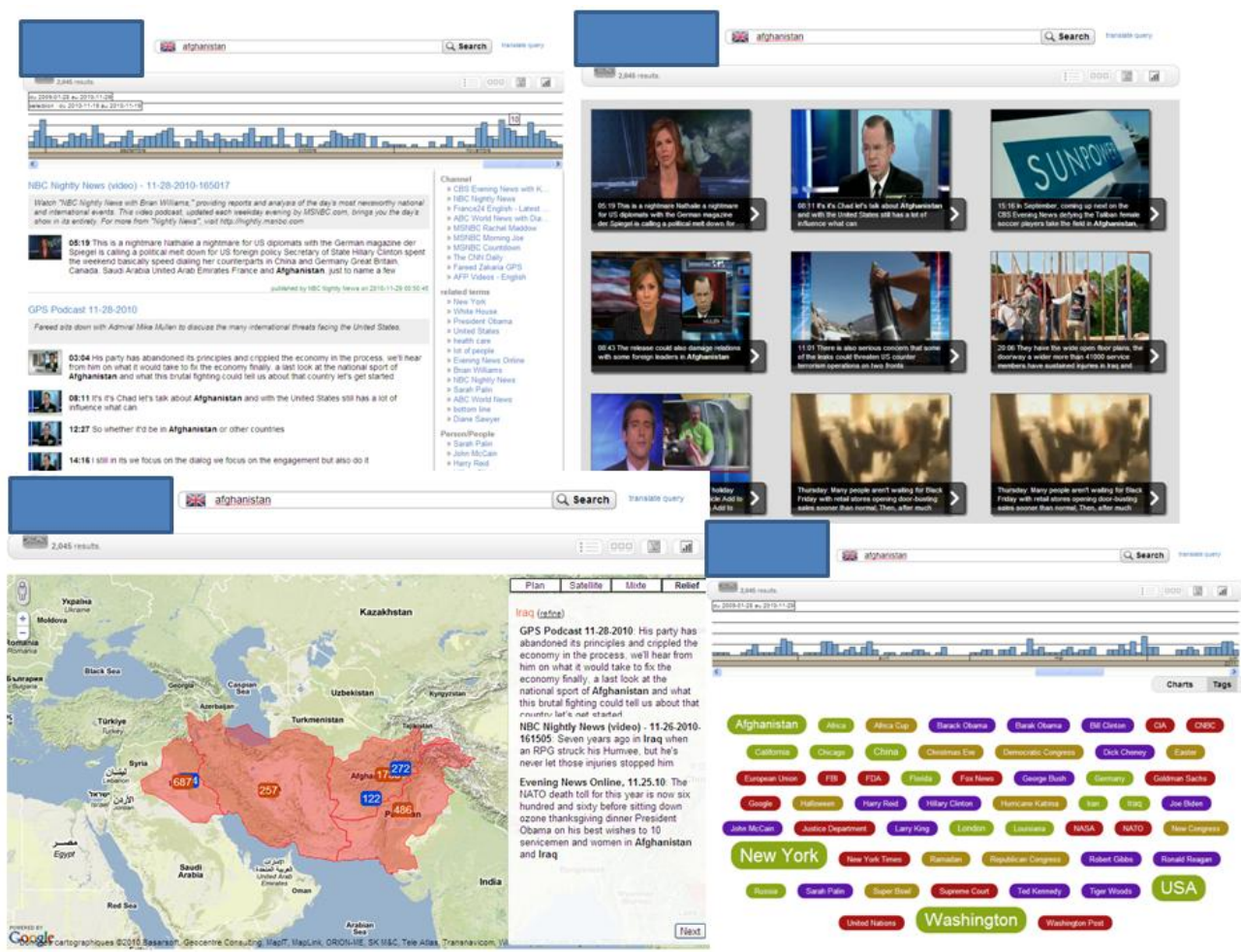

**Figure 3 Different displays of the same query results**

### 2.3 Index and semantic filters

Once speech is transcribed into text and segmented, standard natural language processing (NLP) techniques are applied to each segment. Extracted words, named entities (people, location, organization and events), and multi-word terms are then indexed, along with their time codes from the original automatic speech recognition, with a search engine. We use a proprietary NLP framework similar to the GATE and UIMA open source frameworks. It is resilient and scalable and has been built to scale to terabytes of input, typically an order of magnitude greater than the open source framework. It is able to efficiently process billions of words, and handle streams of hundreds of thousands of tokens per seconds. This NLP framework has been tested over several billion web pages.

### 3 BACK END: OFF LINE PROCESSING

Once video and audio streams have been segmented and processed, the user is presented with a welcome screen that shows trending people, places, locations, and events over a user definable period (day, week, months). This welcome screen allows users to keep abreast of the latest news, without formulating a query. The user can also search in the unique search box, as in classical search.

### 3.1 Index and semantic filters

The welcome page gives an overview of current news using tag clouds or chart pies, underneath a familiar search box. It presents named entities (people, location, organization and event) automatically extracted from all podcast news within a specified time period selected by the user on the time line. Figure 2 on the previous page shows two examples: if we select a period in August 2010, we can see that the event that appears the most is Ramadan. In the second example, the period is end of October and the main event is Halloween. Clicking on the word in the tagcloud or on a slice in the piechart executes a query over the broadcasts. A query can also be posed as in any classical search engine, with the user typing keywords in the search box. An autocompletion service is provided, helping to spell named entities known to the system. All advanced search functions are available, as with classic web search engines: matching exact phrases, logical and regular expressions. Lemmatization and the use of stop words are specific to the language selected by the user.

### 3.2 Result page thumbnails

In response to a query, we have provided a rich interface with multiple views on the results. In addition to thumbnails, timelines, and tag clouds views, the user now has access to a map view showing locations mentioned in fetched segments, as well as automatically determined trends in these segments. These maps and trends are all calculated from the ASR transcription of the audio streams of the broadcasts. Search is performed using a navigation look and feel that is familiar to search engine users.

Figure 3 on the previous page illustrates these different views with the query "Afghanistan". The list view is the classic search engine results page, presenting a list of hits with text snippets related to the query. Thumbnails are contextual, drawn from time related to the text snippet. The user can launch the video at the exact time by clicking on the corresponding thumbnail. On the right of this screen, faceted search provides another mechanism for search and refinement using metadata (such as the source of the podcasts) or terms extracted from the broadcast content (such as named entities). Another view on the same data presents the results as large thumbnails with overlayed text snippets. By clicking on the next icon (the white arrow at eh bottom of the thumbnail) the user can scroll through the parts of the video when the query is mentioned, with appropriate thumbnails related to the moments when the words are mentioned. The map view (shown in the lower left of Figure 3) offers the possibility to see the locations mentioned, automatically extracted from the transcribed text. When clicking on the map on Iraq, for example, we can see the contextual snippets of the results set that correspond to the query Afghanistan with a refine on Iraq: both words are highlighted giving a quick overview of the context to the user. The trends view (lower right corner of Figure 3) sometimes provides a better visualization of the facets related to the query using charts or tag cloud.

In all the views, a histogram time-line allows users to restrict the search to a specific time period. It also allows the user to visualize the number of times that the query term has been mentioned over time.

### 3.3 Segment browsing

Once a query result has been selected, the corresponding video is streamed, starting directly from the segment relevant to the query. The timeline of the video player presents markers with snippets of 30 words containing the query-relevant keywords corresponding to moments where the query words are mentioned. As illustrated in Figure 4, where the query was *Tony Curtis*, we see on the timeline that the last segment of this CBS news broadcasts is the targeted segment that deals with Tony Curtis.

On the same page, we can see that other segments are presented. A thin white progression line shows the position of the current play within the whole video. When the user positions the cursor on a segment bar, we show

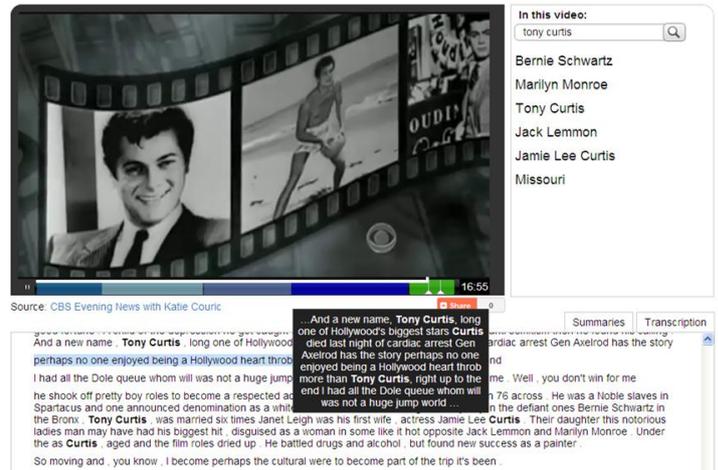

**Figure 4 Video output page with segmentation**

keywords that are mentioned in this segment (shown here as white text in a black box). Associated to these keywords, the named entities that correspond only to the segment are displayed on the right of the video player. These labels and named entities highlight the topics that are discussed in the segment[2].

Figure 5 shows some of the other segments in the same broadcast. In the first segment, extracted keywords shown as white text in black boxes (too small to read here) are "storm", "Carolina" and the named entities related are "Carolina beach", "New Jersey". The report was about a storm on the East Coast of the US, and that the automatically chosen labels were relevant in this case. In the second case the keywords were "foreclosure", "home" and "year"; "year" is obviously too generic but "foreclosure" describes the main subject of that segment.

### 3.4 Cross-language search

Another feature we implemented for searching and browsing videos is cross lingual information retrieval, i.e., querying in one language and see the results from other languages. In figure 6, a query *Afghanistan* is posed in English (as illustrated by the flag). By clicking on the search in other languages, the application uses external translation services to translate the query and display the results in the selected language (Arabic, in the example). The results are displayed in Arabic but can also be translated in English: the snippets and the facets are translated but can still be clicked to play the podcast. The play page can also be translated as a whole, therefore, the search can be in other languages and displayed in any language in an easy and intuitive way for the user.

### 4    PERFORMANCE

Two servers are used in this demonstrator. The first one is for the back-end fetching and indexing, in form of a distributed task scheduler.

---

[2] The named entities extracted from this segment are all related to the query: *Bernie Schwartz* is his real name, *Jamie Lee Curtis* is his daughter and *Marilyn Monroe* and *Jack Lemmon* starred with him in *Some Like It Hot*.

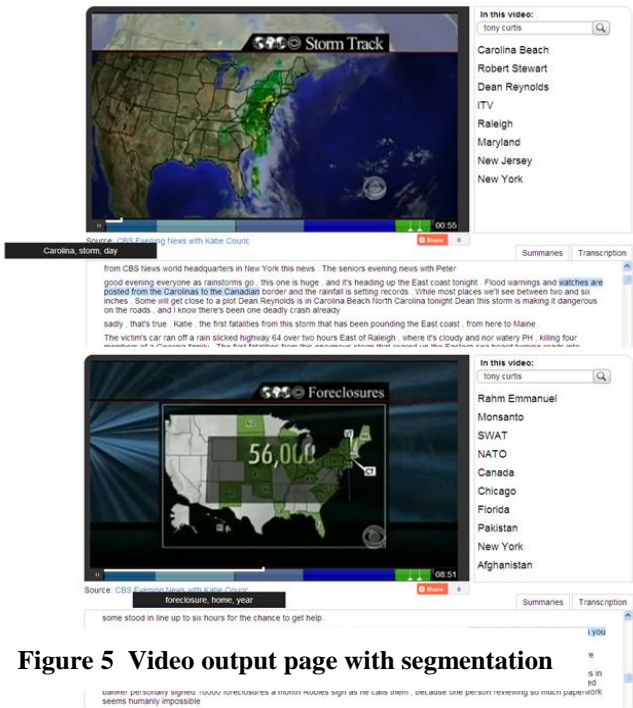

**Figure 5 Video output page with segmentation**

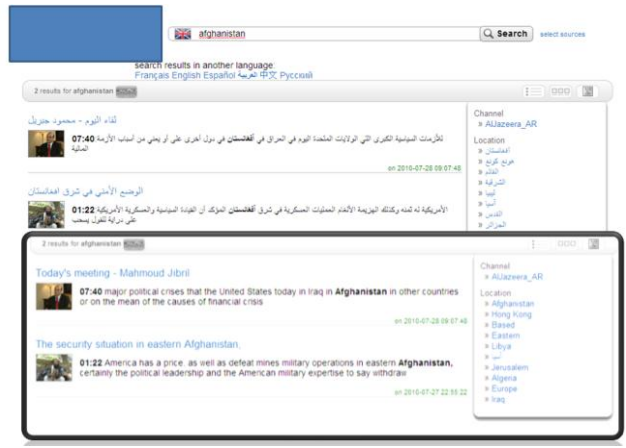

**Figure 6 Cross lingual Search**

Our system is built around a highly robust and distributed task queue, with replaceable workers, as in any map reduce system. Each task is scheduled on a worker (multiple workers per machine) by the queue manager and executed. Workers can be removed or added without any task being lost. The system processes more than 150 new video and audio items each day, amounting to roughly 3.5 Gb or 15 hours of new daily content, on a single 6 core server. Potentially, this server could ingest and process about 100 hours of videos per day. As this backend is fully distributed, servers can be immediately added in the system to handle any load. Another easily parallizable server is dedicated to handling user queries, able to handle 500 unique visitors per day. This configuration is highly scalable and is very similar to our commercial website exalead.com/search (which holds 16 billion web pages). Voxalead currently contains more than 70,000 podcasts.

## 5 CONCLUSION AND NEXT STEPS

Here we described our system for searching and browsing news videos by their language content. We provide many complementary ways to access video and audio content, but the main novelty of our system is that video content is not presented as a single block, but is segmented by its content, and accessible in query-dependent segments. Podcast content is currently derived essentially from the audio part however, the planned next steps are to augment this content via deeper image processing tools like OCR and face recognition. We are also considering adding speaker recognition for the next version.

### Acknowledgments

This research was partially funded by the French funding agency OSEO in the QUAERO project. Vocapia Research provided the ASR (vocapia.com)